\documentclass[a4paper,12pt]{article}
\usepackage{epsfig}
\newcommand{\be}{\begin{equation}}
\newcommand{\ee}{\end{equation}}
\newcommand{\bea}{\begin{eqnarray}}
\newcommand{\eea}{\end{eqnarray}}
\begin{document}

\begin{center}
{\Large \bf Transverse Momentum Distribution of {\boldmath $J/\psi$}'s\\
            in Pb-Pb Collisions and Gluon Rescattering }\\[5mm]
 Pengfei Zhuang$^{[1]}$ and J\"org H\"ufner$^{[2]}$\\[5mm]
 $^{[1]}$ Physics Department, Tsinghua University, Beijing 100084, China 
          \\[2mm]
 $^{[2]}$ Institute for Theoretical Physics, University of Heidelberg,
          D-69120 Heidelberg, Germany\\[5mm]
\end{center}

\begin{abstract}
\setlength{\baselineskip}{16pt}
The transverse momentum dependence of $J/\psi$ suppression
 in relativistic heavy 
ion collisions is calculated from initial state  gluon rescattering
not only  with 
nucleons but also with
 prompt gluons produced in nucleon-nucleon collisions in the early 
phase of the reaction. The second mechanism is new and turns out to be
important in a description of the data for central collisions.
\end{abstract}
\vspace{0.1in} \setlength{\baselineskip}{18pt}
 
\section {Introduction}
Recently, the NA50 collaboration has presented the final analysis of
 transverse  momentum dis\-tri\-bu\-tions\cite{na501} for charmonia
 produced in Pb-Pb collisions  at 158 A GeV. For the $J/\Psi$, the
 values of $\langle p_t\rangle,  \langle p_t^2\rangle$ and $\langle
 M_t - M\rangle$ exhibit a similar trend as  a function of the
 centrality of the collisions: they first increase and then 
 flatten when the centrality increases, and finally turn steeply
 upward for  the most central collisions.

It is commonly accepted\cite{vogt} that the new data on $J/\Psi$
suppression in Pb-Pb collisions
cannot be  explained only by 
inelastic collisions of the $J/\Psi$  with the nucleons of projectile and
target. The anomalous suppression needs a  new absorption mechanism, and
the observed drop at large $E_t$ ($\simeq 100$ GeV)
 is either considered as the expected
onset of $J/\Psi$ melting\cite{blaizot1,matsui} in the picture of the
quark-gluon plasma, or interpreted\cite{blaizot2,capella,hufner2} 
as an effect of
the transverse energy fluctuations around its mean  value determined
by the collision geometry. The data on $J/\Psi$ transverse momentum
distributions provide us additional interesting information and 
possibly  a
deeper insight into the underlying physics.

In a recent publication\cite{hufner1} we have
drawn the attention to a close correlation of the values of $\langle
p^2_t\rangle (E_t)$ and the degree of anomalous suppression $\Delta S
(E_t) =S_G(E_t)-S_{ob}(E_t)$, where $S_G(E_t)$ $(G$=``Glauber'') is
the suppression calculated from the inelastic scattering of the
$J/\psi$ on nucleons alone, while ``$ob$''  refers to the  observed
values. In the same paper we have proposed one explanation for this
correlation: The escape of high $p_t\,J/\psi$ from \underline{final
state} absorption in a QGP or an environment of
comovers. Unfortunately,
the theory required a new parameter. In the present paper we take an
alternative point of view: We investigate whether the data on $\langle
p^2_t \rangle(E_t)$ can be described by an anomalous mechanism in the
\underline{initial state}. We have followed this line, since the bulk
part of the observed $\langle p^2_t\rangle$ seems to be generated in
the initial state.

It is generally agreed that  the additional transverse
momentum broadening $\delta p^2_t=\langle p^2_t\rangle_{AB}-\langle
p^2_t\rangle_{NN}$ observed in nucleus-nucleus (AB) collisions over
the one in nucleon-nucleon (NN) collisions arises in the initial state,
i.e. before the  production of the $c\bar c$ pair\cite{hufner3}.
 The two partons
(mostly gluons) traverse nuclear matter  thereby colliding  with
nucleons (thus acquiring additional $p^2_t$) before they fuse to the
$c\bar c$. Indeed, the data for $\langle p^2_t\rangle_{AB}$ show a
linear relation with the path length of the gluons in the initial
state as  is well studied in $pA$ and $S-U$ collisions\cite{na501}. In a
nucleus-nucleus collision, especially with heavy nuclei, an additional
phenomenon arises, the phenomenon of wounded nucleons. In the approach
of prompt gluons proposed in refs.\cite{hufner4,hufner5}
for the description of anomalous $J/\psi$ suppression ``wounded''
nucleons are nucleons which contain additional prompt bremsstrahl
gluons whose number is proportional to the number of inelastic $NN$
collisions. In the approach of refs.\cite{hufner4,hufner5} the
additional gluons lead to a larger inelastic cross section for a
$J/\psi$  on a wounded nucleon compared to a ground-state nucleon.
 
The wounded nucleon approach should also have an effect on the initial
state. The gluons from projectile and target nuclei, which eventually
fuse to form the $c\bar c$, encounter nucleons and wounded nucleons
before they fuse. The interaction of these gluons with nucleons is
identified as the main source of the additional transverse momentum
broadening $\delta p^2_t$. If the encountered nucleon is wounded,
$\delta p^2_t$ should be even larger. It is this effect, which we
calculate in this paper.

In our calculation we rely heavily on the notation and on the
parameters chosen in refs.\cite{hufner4,hufner5}, yet our presentation
is essentially self-contained. We first present the formulae for
$J/\psi$ suppression in Pb-Pb collisions calculated from inelastic
$J/\psi\ N$ interactions and from collisions with prompt gluons in the
final state. Then we turn to the initial state and derive the formulae
for the generation of the transverse momentum (Sect. 2). Then, in
Sect. 3, we present the numerical results and compare with experiment,
followed by the conclusions (Sect. 4).

\section {Generation of additional transverse 
momentum via interaction with prompt gluons in the initial state}
Fig. 1 is a schematic representation of $J/\psi$ production in
nucleus-nucleus collisions with particular emphasis to the prompt
gluons in the initial and final state.
The collision of a row of projectile nucleons on a row of target
nucleons is depicted in a two-dimensional (time-longitudinal
coordinate) 	plane and in the $NN$ c.m.s.
 A gluon from a projectile nucleon and a gluon from a target
nucleon fuse to produce a premeson at point $O$ which has three
dimensional space coordinates $(\vec s, z_A )$ in the rest frame of
the projectile A and $(\vec b - \vec s,z_B)$ in the rest frame of the
target B, where $\vec b$ is the impact parameter for the AB
collision. The premeson propagates through the projectile and the target,
and develops in time towards the asymptotically observed
$J/\Psi$. During the propagation the premeson interacts with nucleons
from the projectile and target. Since these nucleons have experienced
collisions with other nucleons before their encounter
with the $c\bar c$
premeson, and since they have radiated bremsstrahl ``prompt'' gluons 
(dashed lines),  the $c\bar c$ 
premeson will also interact with these radiated gluons. Taking into
account the two kinds of final state collisions of the premeson 
 the $J/\Psi$
suppression function can be written as \be\label{suppression} S(E_t) =
\int d^2 \vec b d^2 \vec s dz_A dz_B K(\vec b,\vec s,z_A,z_B,E_t)\ ,
\ee where $E_t$ is the transverse energy which reflects the centrality
of the  collision, and the expression for the suppression kernel $K$ is
taken as\cite{hufner5}:
 \be\label{kernel} K(\vec b,\vec s,z_A,z_B,E_t)
= \rho_A(\vec s, z_A)\rho_B(\vec b -\vec s,z_B) e^{-(I_A +
I_B)}P(E_t|b)\ .  \ee 

%%%%%%%%%%%%%%%%%%%%%%%%%%%%%%%%%%%%%%%%%%%%%%%%%%%%%%%%%%%%%%%%%%%%%%%%
\begin{figure}[ht]
\vspace*{+0cm}
\centerline{\epsfxsize=10cm  \epsffile{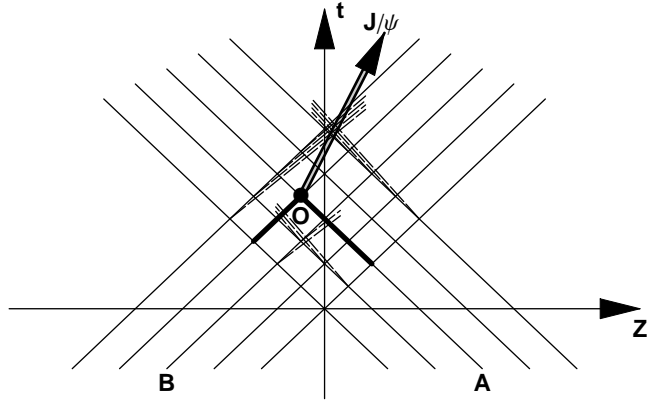}}
\caption{\it A two-dimensional (time-longitudinal coordinate) plot for
$J/\Psi$ production in a collision of nuclei A and B in the c.m.s. of
the colliding nucleons. Nucleons are denoted by thin solid lines, the
two gluons which fuse to form a premeson at point $O$ are denoted by
thick solid lines, while the bremsstrahl gluons are drawn by dashed
lines.}
\label{fig1} 
\end{figure}
%%%%%%%%%%%%%%%%%%%%%%%%%%%%%%%%%%%%%%%%%%%%%%%%%%%%%%%%%%%%%%%%%%%%%%%%

Here $\rho_{A,B}(\vec r)$ is the density of the
nuclei A and B, respectively. The final state $J/\Psi$ absorption via the
interaction with nucleons and prompt gluons in the geometric row
labeled by the variables $\vec b,\vec s$ is contained in the
two exponentials given by \bea\label{iaib} I_A(\vec b,\vec s,z_A,z_B)
&=& \int^\infty_{z_A}dz\sigma_{abs}^{\Psi N}(t_\Psi)\rho_A(\vec
s,z)\nonumber\\ &+&\sigma_{in}^{NN}\int_{z_A}^\infty dz
\int_{-\infty}^\infty dz' \sigma_{abs}^{\Psi
g}(t_\Psi)\Theta(t_g)\langle n_g\rangle (t_g) \rho_A(\vec
s,z)\rho_B(\vec s-\vec b,z')\ ,\nonumber\\ I_B(\vec b,\vec s,z_A,z_B)
&=& \int_{-\infty}^{z_B}dz \sigma_{abs}^{\Psi N}(t_\Psi)\rho_B(\vec s
-\vec b,z_B)\nonumber\\ &+&\sigma_{in}^{NN}\int^{z_B}_{-\infty} dz
\int_{-\infty}^\infty dz' \sigma_{abs}^{\Psi
g}(t_\Psi)\Theta(t_g)\langle n_g\rangle (t_g)\rho_B(\vec s -\vec
b,z)\rho_A(\vec s,z')\ , \eea where $\sigma_{in}^{NN}$ is the
inelastic NN cross section, $\langle n_g\rangle$ is the mean number of
 prompt gluons produced in a NN collision, and $\sigma_{abs}^{\psi
N}$ and $\sigma_{abs}^{\Psi g}$ are the $J/\Psi$ absorption cross
sections for its interaction with a nucleon and with a prompt
gluon, respectively. Effects of formation time for the development of
a $c\bar c$ pair to the $J/\psi$ and for the radiation of a gluon are
taken care of by the  time dependence of $\sigma^{\Psi N}_{abs}
(t_\Psi)$ and $\langle n_g\rangle(t_g)$, respectively. Here  
$t_\Psi$ is the time difference between the $\Psi N$ interaction
point and the formation point $O$ of the premeson, and $t_g$ is the
time difference between the $\Psi g$ interaction point and the $NN$
interaction point which produces the  prompt gluon. The
$\Theta$ functions in $I_A$ and $I_B$ ensure that only the prompt
gluons, which are created before the interaction with the effective
$J/\Psi$, are taken into account.

Using the two-channel model for the evolution of a $c\bar c$ pair, the
time dependent effective cross section $\sigma_{abs}^{\Psi N}$ can be
expressed as\cite{hufner6} \be\label{spsin} \sigma_{abs}^{\Psi
N}(\tau) = \sigma_{in}^{\Psi N} + (\sigma_{in}^{pre} -
\sigma_{in}^{\Psi N})\cos (\Delta M \tau)\ , \ee where
$\sigma_{in}^{pre}$ is the initial 
($\tau\rightarrow 0$)  and $\sigma^{\Psi N}_{in}$ the asymptotic
 ($\tau \rightarrow \infty$)
absorption cross section,  and $\Delta M$ is the mass
difference between $J/\Psi$ and $\Psi'$. We use values
$\sigma_{in}^{\Psi N} = 6.7\ mb$ and $\sigma_{in}^{pre}  = 3\
mb$\cite{hufner5}.

In the constituent quark model, the inelastic $\Psi g$ cross section
differs from        the inelastic $\Psi N$ cross section by a factor
$3/4$\cite{hufner4}, \be\label{spsig} \sigma_{abs}^{\Psi g} = {3\over
4}\sigma_{abs}^{\Psi N}\ .  \ee

Within  pQCD the time dependence of the mean number $\langle n_g\rangle$
of the radiated gluons can be calculated\cite{hufner4} with an
adjustable parameter $\omega_{min}$, the soft limit of the transverse
momentum carried by the prompt gluons. In order to reduce the
numerical efforts, it is parameterized\cite{hufner5} as a step
function  
\be\label{ng} \langle n_g\rangle (t_g)=\left\{\begin{array}{ll} n_g^0\
{t_g\over t_0},\ \  & t_g < t_0\\ n_g^0,\ \ & t_g>t_0 \ .
\end{array}\right.  \ee
 In our numerical calculations  we have used fixed
values\cite{hufner5} $t_0 = 0.6\ fm/c$ and $n_g^0 = 0.75$
(corresponding to $\omega_{min} = 0.75$ GeV) for Pb-Pb collisions at
158 A GeV.

To include the effect of transverse energy fluctuations which have been
shown significant for the  explanation of  the sharp $J/\Psi$
suppression in the domain of very large $E_t$ values, we relate the
mean number $\langle n_g\rangle$ to the transverse energy in the same
way as the energy density\cite{blaizot2} and the comover
density\cite{capella}, \be\label{nget} \langle n_g\rangle \rightarrow
\langle n_g\rangle {E_t\over \langle E_t\rangle(b)}\ , \ee where the
mean value $\langle E_t\rangle(b)$ is proportional to the number of
participant nucleons at fixed $b, \langle E_t\rangle(b) = q N_p(b)$,
and $N_p(b)$ is determined by the collision geometry. The function
$P(E_t|b)$ in the suppression kernel (\ref{kernel}) describes the
distribution of transverse energy in events at a given impact
parameter $b$. It is  chosen as a  Gaussian
distribution\cite{blaizot2} \be\label{spsig} P(E_t|b) = {1\over
\sqrt{2\pi q^2 a N_p(b)}} e^{-{(E_t-\langle E_t\rangle(b))^2\over 2q^2 a
N_p(b)}} \ee with the parameters $q=0.274 GeV$ and $a=1.27$\cite{blaizot2}.
   
We now turn to calculate the $J/\Psi$ transverse momentum within the
mechanism of gluon rescattering in the initial state. Before the two
gluons, one  from the projectile and one  from the target, fuse to
form the $c\bar c$ premeson at point $O$, either of them may
 scatter off one or
several nucleons and one or several prompt gluons, as shown in
Fig.1. Via the  exchange of  transverse momentum at each interaction
vertex, the mean squared transverse momentum of the $J/\Psi$ is a
function of the history of the two gluons, \be\label{pt2}  \langle
p_t^2\rangle(\vec b,\vec s,z_A,z_B) = \langle p_t^2\rangle_{NN} +
{\langle p_t^2\rangle_{gN}\over \lambda_{gN}}\ell_{gN}(\vec b,\vec
s,z_A,z_B)+{\langle p_t^2\rangle_{gg}\over
\lambda_{NN}\lambda_{gg}}\ell_{gg}^2(\vec b,\vec s,z_A,z_B)\ . \ee
Here the first term is the contribution from the $c\bar c$ production
vertex in an isolated NN collision, while the second and third terms
arise, respectively, from multiple $gN$ and $gg$ collisions before
fusion. Since the number of the $gN$ collisions is proportional to the
number of nucleons  which the two gluons encounter, the second
term scales with the sum of the lengths of the 
two gluon paths, \be\label{lgn1}
\ell_{gN}(\vec b,\vec s,z_A,z_B) = \int_{-\infty}^{z_A}dz \rho_A(\vec
s,z)/\rho_0 + \int_{z_B}^\infty dz \rho_B(\vec s-\vec b,z)/\rho_0\ .
\ee The constant before the length $\ell_{gN}$ in eq.(\ref{pt2})
depends on the mean transverse momentum $\langle p_t^2\rangle_{gN}$
acquired in a $gN$ collision and on the mean free path $\lambda_{gN}$
of a gluon in nuclear matter. 
The third term in eq.(\ref{pt2}) is proportional to the number of $NN$
collisions in which the  prompt gluons are generated.
Therefore, $\ell_{gg}^2$  is a sum
of two 2-dimensional integrations (similar to the second terms
in $I_A$ and $I_B$ in the final state interaction),  \bea\label{lgn}
\ell_{gg}^2(\vec b,\vec s,z_A,z_B) &=&
\int^\infty_{z_B}dz\int_{-\infty}^{z_A}dz'\langle
n_g\rangle(t_g)\rho_A(\vec s,z)\rho_B(\vec s-\vec
b,z')/\rho_0^2\nonumber\\ &+& \int_{-\infty}^{z_A}dz\int_{z_B}^\infty
dz' \langle n_g\rangle (t_g)\rho_B(\vec s-\vec
b,z)\rho_A(\vec,z')/\rho_0^2\ .  \eea 
The other factors in the last term of eq.(\ref{pt2}) are: The mean
free path $\lambda^{-1}_{NN}=\rho_0\sigma^{in}_{NN}$ for inelastic
$NN$ scatterings (the source of the prompt gluons) and
$\lambda^{-1}_{gg}=\rho_0\sigma_{gg}$ the mean free path for the
interaction of a gluon (which fuses to form the $c\bar c$) with a
prompt gluon. We use
$\sigma_{in}^{NN} = 32\
mb$ and $\rho_0 = 0.17/fm^3$, while $\langle
p_t^2\rangle_{gg}/\lambda_{gg}$ is considered as a free
 parameter.  The effect of
transverse energy fluctuations can also be included in the initial state
interaction by making the transformation (\ref{nget}) for the mean
number $\langle n_g\rangle$ of the prompt gluons, like in the final
state interaction.

We will call $\ell_{gg}^2$ the squared gluon length related to $gg$
interactions.  However, it is necessary to note that while $\ell_{gN}$
is a geometrical length, either
term in $\ell_{gg}^2$ is not a simply product of the two 
geometrical lengths, since  production time effects of the
prompt gluons are included via  the mean number $\langle n_g\rangle$.

The expression (\ref{pt2}) for the 
mean transverse momentum holds for a  $J/\Psi$ produced
from the rows with coordinates $\vec b, \vec s, z_A$ and $z_B$. In order
to compare with the experimental data, we have to average.
 The average of a  quantity $Q(\vec b, \vec s,z_A,z_B)$
 at fixed
transverse energy $E_t$ is calculated from the suppression kernel $K$
(eq.(\ref{kernel}))
\be\label{average} \langle Q\rangle(E_t) = \int d^2\vec b d^2\vec s
dz_A dz_B Q(\vec b,\vec s,z_A,z_B)K(\vec b,\vec s,z_A,z_B,E_t)/S(E_t)\
.  \ee 
Inserting expression eq.(\ref{pt2}) into eq.(\ref{average}) one obtains
 \bea\label{meanpt2} \langle p_t^2\rangle(E_t) &=& \langle
p_t^2\rangle_{NN} + {\langle p_t^2\rangle_{gN}\over
\lambda_{gN}}\langle\ell_{gN}\rangle(E_t)+{\langle
p_t^2\rangle_{gg}\over
\lambda_{NN}\lambda_{gg}}\langle\ell_{gg}^2\rangle(E_t)\nonumber\\
&=&\langle p_t^2\rangle_{NN}+{\langle p_t^2\rangle_{gN}\over
\lambda_{gN}}\langle\ell_{gN}^{eff}\rangle(E_t)\ , \eea where we have
introduced the effective gluon length $\langle \ell_{gN}^{eff}\rangle$ in
nuclear matter defined by \be\label{eff}
\langle\ell_{gN}^{eff}\rangle(E_t) = \langle\ell_{gN}\rangle(E_t)
+{\langle p_t^2\rangle_{gg}/(\lambda_{NN}\lambda_{gg})\over \langle
p_t^2\rangle_{gN}/\lambda_{gN}}\langle \ell_{gg}^2\rangle(E_t)\ .  \ee

\section{Results and discussions}
To simplify the numerical calculations we use a constant nuclear
density $\rho(r)=\rho_0\Theta(R_A - r)$. Since the calculation of the
average transverse momentum needs information of the suppression
function $S(E_t)$, we first calculate the ratio of the $J/\Psi$ to
Drell-Yan cross sections $\sigma_\Psi(E_t)/\sigma_{DY}(E_t) =
S_\Psi(E_t)/S_{DY}(E_t)$. Here $S_\Psi$ is the $J/\Psi$ suppression
function eq.(\ref{suppression}), and the Drell-Yan cross section $S_{DY}$
is obtained by putting $I_A=I_B=0$ in the kernel $K$ of $S$. The
results are shown in Fig.2 as a function of $E_t$ together with the
new data of the NA50 collaboration. The calculation is normalized at $E_t
= 0$ (NN collision) to $\sigma_\Psi/\sigma_{DY} = 50$. In the absence
of the Gaussian distribution in $E_t$ (eq.(\ref{spsig})) and its
fluctuations related to $\langle n_g\rangle$ (eq.(\ref{nget})),
 we have the same results
as in\cite{hufner5}. It is obvious that the final state
interaction with only nucleons (dashed line) cannot reproduce the
data. The results including final state interactions with prompt gluons
(thin solid line) fit the $E_t$-dependence well, except for the fast
drop at very large $E_t$ values. According to
 refs.\cite{blaizot2,capella,hufner3}, the drop at high $E_t$ arises
from  transverse energy fluctuations. Including the $E_t$
fluctuations via eq.(\ref{nget}) the agreement with data is improved,
but less than the data seem to require. The quality of the fit is
about as good as in ref.\cite{capella} and the residual discrepancy in
the agreement with data may have its origin in the $E_t$ assignment
(c.f. ref.\cite{capella}).

%%%%%%%%%%%%%%%%%%%%%%%%%%%%%%%%%%%%%%%%%%%%%%%%%%%%%%%%%%%%%%%%%%%%%%%%
\begin{figure}[ht]
\hspace{+0cm}
\centerline{\epsfxsize=10cm  \epsffile{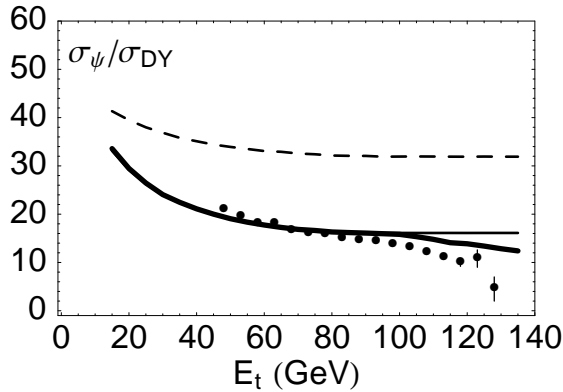}}
\caption{\it The relative $J/\Psi$ production cross section
$\sigma_\Psi/\sigma_{DY}$ in Pb-Pb collisions at 158 A GeV\cite{na503}
as a function of transverse energy. The dashed line corresponds to the
normal nuclear suppression, the thin solid line is obtained with final
state rescattering with nucleons and prompt gluons, and the thick
solid line is calculated including the fluctuations in transverse
energy.}
\label{fig2} 
\end{figure}
%%%%%%%%%%%%%%%%%%%%%%%%%%%%%%%%%%%%%%%%%%%%%%%%%%%%%%%%%%%%%%%%%%%%%%%%

The average gluon lengths (eq.(\ref{eff})) 
are plotted in Fig.3 as a function of
$E_t$. The length $\langle \ell_{gN}\rangle$ in nuclear matter (dashed
line) and the length $\sqrt{\langle\ell_{gg}^2\rangle}$ related to
prompt gluon contribution but without consideration of transverse
energy fluctuations (thin solid line) have the same behavior and
almost the same amplitude. They increase in the domain of small
transverse energies ($E_t < 50 $ GeV), and then  saturate with
$E_t$ for values $E_t > 50$ GeV, where one observes large anomalous
suppression and the prompt gluons become important. The saturation of
the gluon lengths and of the transverse momentum (shown below) in
this domain is due to the saturation of the amount of matter traversed
by the initial gluons in the   Pb-Pb
collisions. As  for the suppression in Fig.2, the contribution of the
transverse energy fluctuations is significant only for very high
transverse energies, $E_t > 100$ GeV (solid line in Fig. 3).

The effects of initial $gN$ and $gg$ scattering to the average
transverse momentum are controlled by the two parameters, $\langle
p_t^2\rangle_{gN}/\lambda_{gN}$ and $\langle
p_t^2\rangle_{gg}/\lambda_{gg}$, respectively. The former may be
extracted from experiments with light projectiles, and the latter is
determined from the fit of eq.(\ref{pt2}) to the data in the $E_t$
region where prompt gluons become important. With the isolated $NN$
contribution $\langle p_t^2\rangle_{NN} = 1.11\ (GeV/c)^2$ and the $gN$
contribution $\langle p_t^2\rangle_{gN}/\lambda_{gN} = 0.081\
(GeV/c)^2/fm$\cite{na501}, the transverse momentum calculated with
only $gN$ scattering (dashed line in Fig.4), namely considering only
the first two terms in eq.(\ref{pt2}), agrees well with the data in
the domain of $E_t < 50$ GeV, while it is clearly below the data
outside this domain. Including the initial scattering with prompt
gluons but without the factor eq.(\ref{nget})
(thin solid line) and choosing  the parameter $\langle
p_t^2\rangle_{gg}/\lambda_{gg} = 1/35\ \langle
p_t^2\rangle_{gN}/\lambda_{gN}$, satisfactory agreement is obtained
for all the data except for the last point, where the transverse
energy fluctuations are expected to play a crucial rule. We include
the fluctuations in $E_t$ via the factor eq.(\ref{nget}) which becomes
relevant only for $\ell^2_{gg}$ in eq.(\ref{pt2}) and arrive at the
solid line in Fig. 4. There is some improvement, but the significance
is difficult to judge, since there is only one data point with rather
large error bar.

%%%%%%%%%%%%%%%%%%%%%%%%%%%%%%%%%%%%%%%%%%%%%%%%%%%%%%%%%%%%%%%%%%%%%%%%
\begin{figure}[ht]
\hspace{+0cm}
\centerline{\epsfxsize=10cm  \epsffile{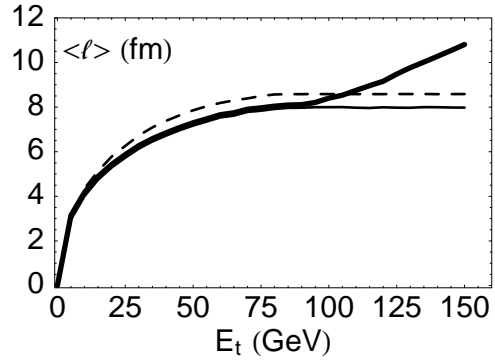}}
\caption{\it The gluon lengths $\langle \ell_{gN}\rangle$ (dashed line)
and $\sqrt{\langle\ell_{gg}^2\rangle}$ without (thin solid line) and
with (thick solid line) fluctuations in transverse energy as a
function of the transverse energy.}
\label{fig3} 
\end{figure}
%%%%%%%%%%%%%%%%%%%%%%%%%%%%%%%%%%%%%%%%%%%%%%%%%%%%%%%%%%%%%%%%%%%%%%%%
%%%%%%%%%%%%%%%%%%%%%%%%%%%%%%%%%%%%%%%%%%%%%%%%%%%%%%%%%%%%%%%%%%%%%%%%
\begin{figure}[ht]
\hspace{+0cm}
\centerline{\epsfxsize=10cm  \epsffile{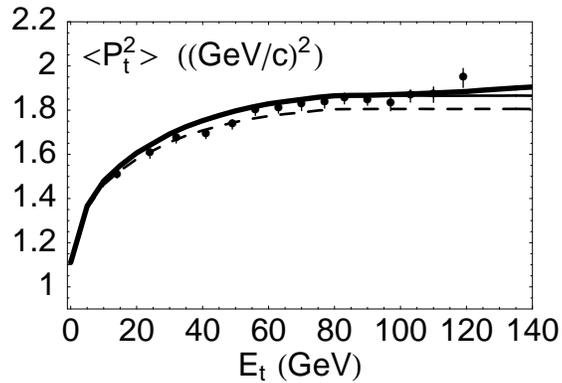}}
\caption{\it The average squared transverse momentum $\langle
p_t^2\rangle$ as a function of transverse energy. The data are 
from\cite{na501}. The dashed line corresponds to the normal gluon
rescattering on nucleons, the thin solid line is obtained with gluon
rescattering with nucleons and prompt gluons, and the thick solid line
is calculated with also the fluctuations in transverse energy.}
\label{fig4} 
\end{figure}
%%%%%%%%%%%%%%%%%%%%%%%%%%%%%%%%%%%%%%%%%%%%%%%%%%%%%%%%%%%%%%%%%%%%%%%%

In order to see the results of our calculations in a different form,
we display the experimental and calculated values for 
$\langle p^2_t\rangle(E_t)$ as a function of
$\langle\ell_{gN}\rangle(E_t)$, Fig. (5a), and as a function of $\langle
\ell^{eff}_{gN}\rangle(E_t)$ (eq.(\ref{eff})), Fig. (5b).
While the dependence $\langle p^2_t\rangle$ versus $E_t$ is
rather complicated with strong saturation effects for large values of
$E_t$, the dependence $\langle p^2_t\rangle$
versus $\langle\ell_{gN}\rangle$ or $\langle\ell_{gN}^{eff}\rangle$ is nearly a
straight line. The dashed and solid lines in Figs. 5a,b correspond to the 
theoretical curves in Fig. 4. According to the definition eq.(\ref{meanpt2}), the slopes 
of the two straight lines are the same and equal to 
$\langle p_t^2\rangle/\lambda_{gN} = 0.081 (GeV/c)^2/fm$.

%%%%%%%%%%%%%%%%%%%%%%%%%%%%%%%%%%%%%%%%%%%%%%%%%%%%%%%%%%%%%%%%%%%%%%%%
\begin{figure}[ht]
\hspace{+0cm}
\centerline{\epsfxsize=10cm  \epsffile{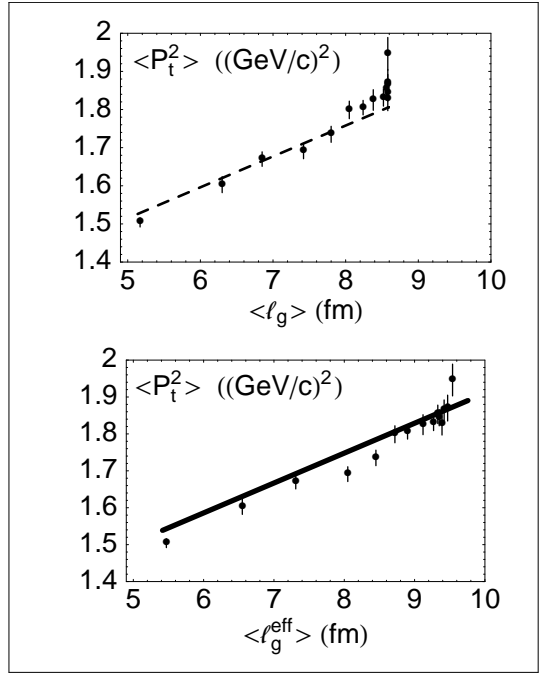}}
\caption{\it The observed $\langle p_t^2\rangle(E_t)$ plotted versus the
calculated $\langle\ell_{gN}\rangle$ and
$\langle\ell_{gN}^{eff}\rangle$. The data are from\cite{na501}.}
\label{fig5} 
\end{figure}
%%%%%%%%%%%%%%%%%%%%%%%%%%%%%%%%%%%%%%%%%%%%%%%%%%%%%%%%%%%%%%%%%%%%%%%%

Until now, we have studied the behaviour  of $\langle p_t^2\rangle$ as
a function of $E_t$. In the experiment under consideration\cite{na501}
the authors have also measured the $p_t$ distributions within a given
$E_t$ interval. They have displayed their data as a ratio
\be\label{ratio} R_{i/1}(p_t) =
{S(p_t|E_t^{(i)})\over S(p_t|E_t^{(1)})} \ee 
where $S(p_t|E^{(i)}_t)$ is the differential distribution in $p_t$,
normalized to $\int d^2 p_t\,S(p_t|E_t^{(i)})=S(E_t^{(i)})$. We have
calculated  $S(p_t|E_t^{(i)})$ by
assuming a Gaussian
$p_t$ dependence. Then the suppression function $S(p_t|E_t^{(i)})$ can be
written as \be\label{spt} S(p_t|E_t^{(i)}) = \int d^2\vec b d^2\vec s
dz_A dz_B K(\vec b,\vec s,z_A,z_B,E_t^{(i)}){1\over \langle
p_t^2\rangle(\vec b,\vec s,z_A,z_B)}e^{-{p_t^2\over \langle
p_t^2\rangle(\vec b,\vec s,z_A,z_B)}}\ . \ee 
For the 
comparison with the experimental data, we divide the $E_t$ region into
five bins with mean values $E_t = 23,\ 48,\ 70,\ 90,\ 110.$ The
theoretical results together with the data are shown in Fig.6.

%%%%%%%%%%%%%%%%%%%%%%%%%%%%%%%%%%%%%%%%%%%%%%%%%%%%%%%%%%%%%%%%%%%%%%%%
\begin{figure}[ht]
\hspace{+0cm}
\centerline{\epsfxsize=10cm  \epsffile{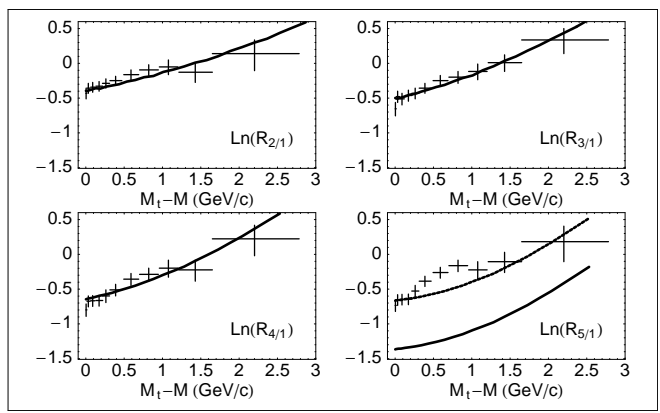}}
\caption{\it The ratio $R_{i/1}$
of the $J/\Psi$ cross section with large value
$E_t^{(i)}$ in the $E_t$ bin $i$ to the cross section with small
$E_t^{(1)}$ in the first $E_t$ bin as a function of $M_t-M$.
We have plotted the logarithm of this quantity, since the global data
seem to be exponentials $\exp(-(M_t-M)/T)$.
 There are five $E_t$ bins corresponding to mean $E_t = 23,\
48,\ 70,\ 90,\ 110\ GeV$. The data are from\cite{na501}. The
theoretical   lines are calculated with initial and final state
rescattering with nucleons and prompt gluons and with transverse
energy fluctuations. For $R_{5/1}$, the solid line is our calculation,
while the dotted line represents the calculated curve normalized to
the data for small $M_t-M$.}
\label{fig6} 
\end{figure}
%%%%%%%%%%%%%%%%%%%%%%%%%%%%%%%%%%%%%%%%%%%%%%%%%%%%%%%%%%%%%%%%%%%%%%%%

For the figures we have not followed ref.\cite{na501}, who plot
$R_{i/1}$ versus $p_t$, but we rather  plot $\ln R_{i/1}$ versus
$\Delta M=\sqrt{M^2+p^2_t}-M$. This representation leads to a straight line
for the data integrated over $E_t$ (cf.\cite{na501}). According to
Figs. 6, this representation also leads to nearly straight lines.  The
solid lines represent our calculations. For all ratios, except
$R_{5/1}$, the data are well accounted for. We see a problem in the
quantity $R_{5/1}$. Our prediction is consistently below the data,
although our integrated value $S(E_t^{(5)})$ with $E_t^{(5)}=110$ GeV
lies \underline{above} the data (Fig. 2). Thus we would have expected
our calculation to be consistently above. In order to see whether
calculated and experimental slopes agree we have normalized the
calculated curve at small $p_t$ to the data. The renormalized curve is
shown by the dots and describes the data reasonably well, though there
may be systematic deviations at large $p_t$. But since the error bars
are too large, speculations about the discrepancies may be dangerous.

\section{Conclusions}  

In this paper, we have studied the transverse momentum dependence of
$J/\psi$  production  in Pb-Pb collisions at 158 AGeV. We have
investigated whether the mechanism of prompt gluons can account for
both effects, anomalous suppression of $J/\psi$ and anomalous values
of $\langle p^2_t\rangle$. Here, ``anomalous'' is judged with respect
to the ``conventional'' values of suppression and $\langle
p^2_t\rangle$ derived from the data of $pA$ and $S-U$ collisions. The
conventional physics is shown as  dashed lines in Figs. 2 and 4,
while the anomaly  is the difference between the data and these
lines. While the anomaly is large (an effect of a factor 2) in the
suppression data, Fig. 2, the anomaly is small for the values of
$\langle p_t^2\rangle$. The difference between the data  and the
dashed line is of order 10\%, if we use $\langle
p_t^2\rangle_{AB}-\langle p^2_t\rangle_{NN}$ as the baseline. Within
the model of prompt gluons we are able to describe the data for the
$J/\psi$ \underline{suppression}
 (except possibly at very high values of $E_t$) by
introducing one adjustable parameter
\be\label{17}
\frac{n^0_g\cdot\sigma^{J/\psi\, g}_{abs}}{\sigma^{J/\psi\, N}_{abs}}=0.56
\ee
where $n^0_g$ is the number of produced  hard prompt gluons per $NN$
collision  (eq.(\ref{ng}))
and the cross sections $\sigma^{J/\psi\, g}_{abs}$ and
$\sigma^{J/\psi\, N}_{abs}$ describe the inelastic cross sections
leading to the destruction of the $J/\psi$ in collisions with a gluon
and a nucleon, respectively. Within the same model we are also able to
describe the \underline{data for $\langle p^2_t\rangle$}, Fig. 4, after a new
parameter
\be\label{18}
\frac{\langle p^2_t\rangle_{gg}}{\lambda_{gg}}\cdot\left(\frac{\langle
p^2_t\rangle_{gN}}{\lambda_{gN}}\right)^{-1}
=1/35\ee
is adjusted. The magnitude  of the values eqs.(\ref{17}) and (\ref{18})
for the two parameters
reflects the observation that the anomaly in the suppression is large
(50\%) while it is a small effect for the values of $\langle
p^2_t\rangle$. 

We conclude: We have presented two possible explanations for the
anomalous behaviour of the dependence of $\langle p^2_t\rangle(E_t)$
in Pb-Pb collisions. In ref.\cite{hufner1} the effect is described as
one of the final state while in this paper the gluon rescattering on
bremsstrahl gluons in the initial state has been investigated as a
possible source. In both approaches, satisfactory agreement with the
data is achieved after adjusting one parameter in each  calculation.

\vspace{0.3in}

\noindent {\bf \underline{Acknowledgments:}} We thank Boris
Kopeliovich who drew our attention to the initial state interactions.
 One of  the authors
(P.Z.) is grateful for the hospitality at the Institute for
Theoretical Physics in  Heidelberg, his work was supported by the
grants 06HD954, NSFC19925519 and G2000077407.

\end{document}